# Inference with interference between units in an fMRI experiment of motor inhibition

Xi Luo, Dylan S. Small, Chiang-shan R. Li, Paul R. Rosenbaum[1]

University of Pennsylvania, Philadelphia

Abstract. An experimental unit is an opportunity to randomly apply or withhold a treatment. There is interference between units if the application of the treatment to one unit may also affect other units. In cognitive neuroscience, a common form of experiment presents a sequence of stimuli or requests for cognitive activity at random to each experimental subject and measures biological aspects of brain activity that follow these requests. Each subject is then many experimental units, and interference between units within an experimental subject is likely, in part because the stimuli follow one another quickly and in part because human subjects learn or become experienced or primed or bored as the experiment proceeds. We use a recent fMRI experiment concerned with the inhibition of motor activity to illustrate and further develop recently proposed methodology for inference in the presence of interference. A simulation evaluates the power of competing procedures.

---

[1] *Address for correspondence:* Xi Luo is assistant professor in the Department of Biostatistics at Brown University. Dylan S. Small is associate professor and Paul R. Rosenbaum is professor in the Department of Statistics, The Wharton School, University of Pennsylvania, 473 Jon M. Huntsman Hall, 3730 Walnut Street, Philadelphia, PA 19104-6340 US. Chiang-Shan R. Li is associate professor of psychiatry and neurobiology, Department of Psychiatry, Yale University, CMHC S103, 34 Park Street, New Haven, CT 06519 US. This work was supported by grants from the Measurement, Methodology and Statistics Program of the U.S. National Science Foundation. Imaging data were collected under the support of NIH grants R01DA023248, K02DA026990 and R21AA018004. 11 August 2011 E-mail: xluo@stat.brown.edu.

# 1  Introduction: An Application of Inference with Interference

## 1.1  What is interference between units?

If treatment effects are defined as comparisons of the two potential responses that an experimental unit would exhibit under treatment or under control (Neyman 1923, Welch 1937, Rubin 1974, Lindquist and Sobel 2011), then an implicit premise of this definition is "no interference between units," as discussed by Cox (1958, p. 19): "There is no 'interference' between different units if the observation on one unit [is] unaffected by the particular assignment of treatments to the other units;" see also Rubin (1986). For instance, widespread use of a vaccine may benefit unvaccinated individuals because they are less likely to encounter an infected individual, a form of interference known as herd immunity; see Hudgens and Halloran (2008). In agriculture, the treatment applied to one plot may also affect adjacent plots; see David and Kempton (1996). In social experiments, people talk, and changing the treatment applied to one person may change what she says to someone else, altering his response to treatment; see Sobel (2006).

In some contexts, interference is of central interest in itself — this can be true of herd immunity or of social interaction, for example — but in many if not most contexts, interference is principally an inconvenience, depriving us of both independent observations and a familiar definition of treatment effects. We apply and extend a recent, general approach to inference with interference (Rosenbaum 2007a) in the context of a cognitive neuroscience experiment in which the brains of a moderate number of subjects are studied using fMRI while faced with a rapid fire sequence of randomized stimuli. In this context, interference is likely to be widespread and difficult to model with precision. The goal is a simple, sturdy, valid method of inference whose conclusions about the magnitude of treatment effects are intelligible when the interference may be complex in form.

## 1.2 Three themes: randomization inference, confidence intervals with interference, ineffective trials

In this case-study, we reanalyze a randomized experiment in cognitive neuroscience with a view to illustrating three ideas, one very old idea, one somewhat new idea, and one idea that has evolved gradually over more than half a century. In many cognitive neuroscience experiments, a moderate number of subjects are repeatedly exposed to many randomly selected stimuli intended to elicit cognitive activity of a specific type together with its characteristic neurological activity visible with, say, fMRI. Three dilemmas arise in these experiments. First, because a few thoughtful, complex human subjects are observed many times performing simple repetitive tasks, subjects become familiar with the tasks, perhaps increasingly bored or skillful or distracted or fatigued or aware of the purpose of the experiment, so the situation is unlike a study of a single response elicited from each of many separated, unrelated subjects, and also unlike a stationary time series or a repeated measures model with dependence within subjects represented by additive subject parameters. In such a context, one might wish to draw inferences about treatment effects on many brain regions without relying on a model fitted to just a few people. Second, for reasons both biological and cognitive, rapid-fire stimuli are likely to interfere with one another, in part because the neurological response to one stimulus is expected to last well beyond the presentation of the next stimulus, and in part because learning and boredom and surprise are global cognitive responses to long segments of a sequence of stimuli not responses to a single stimulus. If 100 treatment/control tasks are presented to one subject in ten minutes, then it is unrealistic to characterize the effect of the treatment versus control in terms of response to single trials, because the response to each trial is affected by many previous trials. We need to characterize the differing responses to treated and control stimuli without assuming the mind and brain are born anew after each stimulus. Third, the experimenter controls the stimuli, the requests for cognitive activity, but requests for cognitive activity may not produce the requested activity, and hence not produce the neurological activity characteristic of that cognitive activity. This is familiar from conversation: a speaker asks a question only to receive the reply: "Would you repeat that? I wasn't listening." If a statistical test is used that expects every stimulus to elicit its intended cognitive activity, the test may have much less power to detect actual activity than a test which acknowledges distraction and boredom and error in addition to the requested activity.

The old idea, due to Sir Ronald Fisher (1935), is that randomization can form the "reasoned basis for inference" in randomized experiments, creating without modeling assumptions all of the probability distributions needed to test the null hypothesis of no treatment effect. In his introduction of randomized experimentation, Fisher (1935, Chapter 2) pointedly used a single-subject randomized experiment — the famed lady tasting tea — precisely because modeling and sampling assumptions seemed so inadequate to describe a single-subject experiment. In particular, there was no need to model the lady's evolving cognitive activity to test the null hypothesis that she could not discern whether milk or tea had first been added to the cup. Although it sometimes receives less emphasis in the statistics curriculum of 2011, Fisher's theory of randomization inference was viewed as one of the field's celebrated results. This method of using random assignment to replace modelling assumptions was described by Jerzy Neyman (1942, p. 311) as "a very brilliant one due to Fisher," and in retrospect Neyman (1967, p. 1459) wrote: "Without randomization there is no guarantee that the experimental data will be free from a bias that no test of significance can detect." In a similar vein, John Tukey (1986, p. 72) recommended: "using randomization to ensure validity — leaving to assumptions the task of helping with stringency." (Stringency is decent power in difficult situations, in the spirit of the formal notion of a most stringent test which minimizes the maximum power loss over a class of alternatives.)

The newer idea addresses a limitation of Fisher's method when used in the presence of interference. Fisher's method yields a valid test of the null hypothesis of no effect. If a treatment effect has a simple form, say an additive constant effect or shift, then it is possible to invert Fisher's test of no effect to yield a confidence statement for the magnitude of this constant effect (e.g., Lehmann 1975); however, by its nature, interference precludes such a simple form for an effect. The newer idea is to invert the randomization test of no effect to yield a confidence interval for an attributable effect in the presence of interference that contrasts the results seen with an active treatment to the results that would have been seen in an experiment of identical design but with no active treatment, a so-called "uniformity trial" common in the early years of randomized agricultural experimentation (Rosenbaum 2007a). This newer idea is applicable with distribution-free statistics whose distribution in the uniformity trial is known without conducting the uniformity trial. The classes of distribution-free statistics and of rank statistics overlap substantially but are not the same, and it is the distribution-free property that is needed here.

The third, gradually evolving idea made a first appearance in a paper by Lehmann (1953) concerned with the power of rank tests. After showing that Wilcoxon's test was the locally most powerful rank test for a constant, additive effect in the presence of logistic errors, Lehmann went on to show that it was also locally most powerful against a very specific mixture alternative in which only a fraction of subjects respond to treatment. Conover and Salsburg (1988) generalized the mixture alternative and derived the form of the corresponding locally most powerful test; this was no longer Wilcoxon's test, but rather a test that gave greater emphasis to larger responses. Although they substantially increase power when some trials fail to elicit the intended effect, the ranks used by Conover and Salsburg have no obvious interpretation, so they cannot be used as the basis for an attributable effect. It turns out, however, that Conover and Salsburg's ranks are almost the same as ranks proposed by Stephenson (1981); see also related work by Deshpande and Kochar (1980) and Stephenson and Gosh (1985). Using Stephenson's ranks, a confidence interval for attributable effects becomes available (Rosenbaum 2007b), thereby permitting inference about the magnitude of the effect in the presence of interference.

Although the main goal is to illustrate these three ideas in the context of an fMRI experiment, along the way a technical issue arises. The experiment is not a perfectly balanced design, and for unbalanced designs the attributable effect has a more natural interpretation if it is not formulated in terms of a linear rank statistic, but rather in terms of a linear placement statistic in the sense of Orban and Wolfe (1982), which is a form of nonlinear rank statistic. Because perfect balance is difficult to achieve in cognitive neuroscience experiments, we develop the formalities in terms of the placement statistics that are most likely to be useful in practice.

### 1.3 A randomized experiment in the cognitive neuroscience of motor inhibition

In the experiment by Duann, Ide, Luo and Li (2009), each of 58 experimental subjects was observed in four experimental sessions that were each about ten minutes in length. At random times during a session, a trial began, with a median of 97 trials per session. With probability $\frac{3}{4}$, the trial was a "go trial:" a dot was presented on a screen, and after a interval of time of random length, the dot became a circle signifying that the subject was to quickly press a button. With probability $\frac{1}{4}$, the trial was a "stop trial:" the trial began as a go trial, but briefly after the circle appeared it was replaced by an X signifying "do not press the button." In a stop trial, the subject is instructed to do something, and then the instruction is cancelled. Here, an experimental unit is a trial, with stop trials called

'treatment' and go trials called 'control.' During an experimental session, brain activity was recorded using fMRI at two second intervals. The experiment sought to determine how the brain reacted differently to go and stop trials, where stop trials call for inhibition of a previously requested motor response from the subject.

Figure 1 shows one session for one subject. The vertical grey lines are go trials. The vertical black lines are stop trials. Based on fMRI, Figure 1 shows activity in the subthalamic nucleus (STN). Aron and Poldrack (2006) and Li et al. (2008) suggested that the STN plays an important role in response inhibition. In the lower portion of Figure 1, the STN activity is filtered without use of the stop/go distinction. The filter is a high-pass filter of 128s: it removes slow, low frequency drifts, leaving behind the high frequency ups and downs thought to reflect brain activity. The effects of the filter are somewhat visible in Figure 1. For the STN, we analyze unfiltered and filtered data in parallel, obtaining similar conclusions. In effect, the experiment produces $58 \times 4 = 232$ figures analogous to Figure 1, one for each subject in each session, and does this for many regions of the brain.

The assumption of "no interference between units" is not at all plausible in Figure 1. A typical session has about a hundred trials or experimental units in about 600 seconds. There is interference if the response of a subject at a given trial is affected by treatments at other trials. Interference is likely for at least two reasons. First, the brain has a measurable response to a stimulus for many seconds after the stimulus has been withdrawn, so a subject is still responding to one trial when the next trial begins. Second, the response to a stop trial preceded by a long string of go trials may be different from the response to a stop trial preceded by another stop trial. In addition, as time goes by, subjects are experiencing the normal responses people have when performing a repetitive task: they become familiar with the task, or bored by the task, or less distracted by the recording equipment and more focused on the task, or distracted by something else, and each of these is a response to their entire past experience. The 22,440 trials in this experiment are nothing like a randomized clinical trial with 22,440 unrelated people who do not interfere with each other. It is, nonetheless, a randomized trial and randomization can form the basis for inference, as it did in Fisher's (1935, §2) prototype trial of one lady tasting eight cups of tea.

The trial has a second important feature. Not all trials are "successful." In the first instance, in a stop trial, the subject is instructed first to "go" — press the button — and then the instruction is cancelled. In a stop trial, if the random time between the circle and the X is longer than usual and the subject is quicker than most, then she may press the button before the instruction is cancelled. In this case, even though the trial is randomized to be a stop or treated trial, her brain should exhibit the response typical under the control condition, because nothing she experienced distinguished the trial from a go trial. In addition to the situation just described, it may also happen that the subject is unambiguously told to press the button but does not do so, or is unambiguously told not to press the button but does so anyway, perhaps because the subject is momentarily distracted. Also, a subject may exhibit correct behavior with erroneous thoughts, say failing to press the button because of distraction or fatigue rather than inhibition. Expressed differently, whether or not a trial is successful is not generally a visible property of the trial, yet we are confident that human subjects do not always think the thoughts an experimenter requests. If a trial is not successful in any of these senses, then the requested cognitive activity may not take place, so there may not be the change in blood oxygenation that would typically accompany the requested cognitive activity. Although a stimulus asks for a cognition, we cannot tell whether the cognition took place or not, because we see only behavior and neurological response, but it is unlikely that every stimulus elicits its intended cognition. We might think of responses as a mixture of successful and unsuccessful trials, where successful trials produce a specific pattern of fMRI response. Salsburg (1986), Conover and Salsburg (1988) and Rosenbaum (2007b) consider rank tests that are particularly effective when only a subset of experimental units respond to treatment. These rank tests score the ranks in such a way that little weight is given to lower ranks. In the current paper, a similar approach is taken in studies with interference between units.

When a region of the brain is stimulated to activity, the change in blood oxygenation measured by fMRI is not immediate. There is a brief delay, perhaps a dip, for about 2 seconds, followed by a sharp rise, a sharp fall to slightly below baseline, followed by a gradual return to baseline; see Lindquist (2008, Figure 3). This curve is known as the hemodynamic response function (HRF). We use the form developed by Friston et al. (1998), specifically a weighted difference of two gamma densities, $\gamma(x; \omega, \vartheta) = \vartheta^\omega x^{\omega-1} \exp(-\vartheta x) / \Gamma(\omega)$, both with parameter $\vartheta = 1/16$, and with shape parameters $\omega_1 = 6$ and $\omega_2 = 16$, specifically the function $\mathrm{hrf}(x) = \gamma(16x; 6, 1/16) - \gamma(16x; 16, 1/16)/6$ where $x$ is in seconds. Although we do not report these results, we tried a second form for the HRF with a similar shape but built from inverse logit functions (Lindquist and Wager 2007), obtaining qualitatively similar results in a table parallel to Table 2.

Recall that the measurements in Figure 1 occur at two second intervals. Evaluating the hemodynamic response function, $\mathrm{hrf}(x)$, at two-second intervals, we computed 17 weights for 17 two-second intervals that follow each trial, that is, for the $2 \times 17 = 34$ seconds that follow a trial. These weights sum to one. The first weight is zero, the third and fourth weights are the largest (.375 and .385), and beginning at the eighth the weights turn slightly negative (weight eight is $-0.031$) gradually returning to zero (weight 17 is $-0.0001$). At the end of a session, if fewer than 17 two-second intervals remained, we used the remaining intervals and renormalized the weights so that they again summed to one. For a region such as STN in Figure 1, after each trial, we computed the sum of the HRF weights multiplied by the fMRI measurements. If a region of a subject's brain has become unusually active, we expect this weighted average to become unusually large. Figure 2 is a pair of boxplots of this weighted average for the one session and one subject in Figure 1. In Figure 2, there is some indication that the responses in the stop or treated trials in Figure 1 are elevated.

Although the analysis uses the responses in Figure 2 weighted by the hrf $(x)$ function, the method is applicable with any method of scoring the trials that produces one number per trial. For instance, a response that is sometimes used is the correlation between the hrf $(x)$ function and the sequence of responses that immediately follow a trial.

## 1.4 Outline

Section 2 reviews notation from Rosenbaum (2007a) for treatment effects when interference may be present. In §3.1, a nonlinear rank statistic $T_\mathbf{Z}$ is proposed for a randomized block design with blocks of unequal block sizes; in particular, $T_\mathbf{Z}$ is intended to perform well when not all treated trials are successful in eliciting the intended cognitive activity. Under the null hypothesis of no treatment effect there is, of necessity, no interference among the treatment effects, and §3.2 uses ideas from Orban and Wolfe (1982) to obtain the null randomization distribution of the test statistic $T_\mathbf{Z}$. A confidence statement about the magnitude of effect with interference is then obtained by a pivotal argument in §3.3: it measures the magnitude of the difference between the actual trial and the uniformity trial. In §4, the method of §3.3 is applied to activation of the subthalamic nucleus in §1.3. A simulation in §5 evaluates the power of $T_\mathbf{Z}$ in experiments with interference using a mixture model in which not all trials elicit the intended cognitive activity.

## 2 Notation: The Randomized Trial and the Uniformity Trial

### 2.1 Blocked randomized trial with interference between units

There are $B \geq 1$ blocks, $b = 1, \ldots, B$, and $N_b \geq 2$ units $bi$ in block $b$, $i = 1, \ldots, N_b$, with $N = \sum N_b$ units in total. In §1, there are $B = 58 \times 4 = 232$ blocks consisting of the four sessions for each of 58 subjects, $N = 22,440$ units or trials in total, with $87 \leq N_b \leq 104$ and a median $N_b$ of 97. In block $b$, $n_b$ units are picked at random for treatment, $1 \leq n_b < N_b$, the remaining $m_b = N_b - n_b \geq 1$ units receive control. In §1, the probability that a unit was a "stop trial" was $\frac{1}{4}$, resulting in $13 \leq n_b \leq 37$ and a median $n_b$ of 24. If unit $i$ in block $b$ was assigned to treatment, write $Z_{bi} = 1$, and if this unit was assigned to control write $Z_{bi} = 0$, and let $\mathbf{Z} = (Z_{11}, Z_{12}, \ldots, Z_{B,N_B})^T$ be the $N$-dimensional vector in the lexical order. For a finite set $A$, write $|A|$ for the number of elements of $A$. Write $\Omega$ for the set containing the $|\Omega| = \prod_{b=1}^{B} \binom{N_b}{n_b}$ possible values $\mathbf{z}$ of $\mathbf{Z}$, so $\mathbf{z} \in \Omega$ if and only if $z_{bi} = 0$ or $z_{bi} = 1$ and $n_b = \sum_{i=1}^{N_b} z_{bi}$ for $b = 1, \ldots, B$. Write $\mathbf{n} = (n_1, \ldots, n_B)^T$ and $\mathbf{m} = (m_1, \ldots, m_B)^T$.

Write $r_{bi\mathbf{z}}$ for the response that the $i$th unit in block $b$ would have if the treatment assignment $\mathbf{Z}$ equalled $\mathbf{z}$ for $\mathbf{z} \in \Omega$. In §1.3, for trial $i$ of subject/session $b$, the response $r_{bi\mathbf{z}}$ is the HRF weighting of either the unfiltered or filtered activity in the subthalamic nucleus (STN). Each unit has $|\Omega|$ potential responses, only one of which is observed, namely $r_{bi\mathbf{Z}}$. Figure 2 plots $r_{bi\mathbf{z}}$ for $Z_{bi} = 0$ and $Z_{bi} = 1$ for one $b$ and $i = 1, \ldots, N_b$.

Unlike the notation of Neyman (1923) and Rubin (1974), the response of the $i$th unit in block $b$ may depend on the treatments $\mathbf{Z}$ assigned to all the units; that is, this notation permits interference (Rosenbaum 2007a). In §1, it is quite plausible that a previous treatment for one subject may affect later responses of this same subject. Indeed, it is possible that interference extends across the four blocks or sessions for a given subject.

Write $\mathcal{R}$ for the unobservable array with $N$ rows and $|\Omega|$ columns having entries $r_{bi\mathbf{z}}$. The unobservable $\mathcal{R}$ describes what would happen under all possible treatment assignments $\mathbf{z} \in \Omega$, but $\mathcal{R}$ does not change when actual randomized treatment assignment $\mathbf{Z}$ is selected. In contrast, the observable responses $r_{bi\mathbf{Z}}$ are one column of $\mathcal{R}$, and which one column that is does, of course, depend upon the randomized treatment assignment, $\mathbf{Z}$. Fisher's (1935) sharp null hypothesis $H_0$ of no treatment effect asserts that $r_{bi\mathbf{z}} = r_{bi\mathbf{z}'}$ for all $\mathbf{z}, \mathbf{z}' \in \Omega$ and all $b, i$, so within each row $bi$ of $\mathcal{R}$ all $|\Omega|$ columns have the same value for $r_{bi\mathbf{z}}$.

No interference between units means that $r_{bi\mathbf{z}} = r_{bi\mathbf{z}'}$ whenever $z_{bi} = z'_{bi}$, that is, the response in block $b$ at trial $i$ depends on the treatment $z_{bi}$ assigned in block $b$ at trial $i$, but it does not depend on the treatments $\mathbf{z}$ assigned at other trials. As discussed in §1.3, no interference between units is not plausible in Figure 1, and because of the overlapping of HRF functions is virtually impossible in Figure 2 if $H_0$ is false.

By a randomized block experiment, we mean that

$$\Pr\left(\mathbf{Z} = \mathbf{z} \mid \mathcal{R},\, \mathbf{n},\, \mathbf{m}\right) = \frac{1}{|\Omega|} \text{ for each } \mathbf{z} \in \Omega. \tag{1}$$

In §1, the timing of trials and hence also the number $N_b$ of trials was determined by a random process; then, with probability $\frac{1}{4}$ the trial was a "stop trial" and with probability $\frac{3}{4}$ the trial was a "go trial;" hence, $\mathbf{n}$ and $\mathbf{m}$ were random variables, but the conditional probabilities given $\mathcal{R}$, $\mathbf{n}$, $\mathbf{m}$ of particular patterns of stop or go trials was completely randomized within each block in the sense that (1) was true. Importantly, (1) says treatment assignments were determined by a truly randomized mechanism that ensured the unobservable potential responses $\mathcal{R}$ were not predictive of treatment assignment $\mathbf{Z}$; this is, of course, the essential element of randomized treatment assignment. In a randomized experiment, any association between treatment assignment $\mathbf{Z}$ and the observed responses, $r_{bi\mathbf{Z}}$, is due to an effect caused by the treatment expressed in $\mathcal{R}$; because of (1), an association between $\mathbf{Z}$ and $r_{bi\mathbf{Z}}$ cannot result from biased selection into treated and control groups if the treatment has no effect, that is if Fisher's $H_0$ is true.

## 2.2 The uniformity trial

As mentioned previously, in the absence of interference, it is natural to ask how a unit would have responded if that one unit had received the treatment and or if that one unit had received the control and to define effect of the treatment on this unit as a comparison of these two potential responses; see Neyman (1923), Welch (1937) and Rubin (1974) for discussion of this standard way of defining treatment effects in randomized experiments. This formulation does not work when there is interference between units because a unit may be affected by treatments applied to or withheld from other units. Some definition of the treatment effect with interference is needed if a randomization test of the null hypothesis of no effect is to be inverted to obtain a confidence interval for the magnitude of effect. In principle, the treatment effect is characterized by the $N \times |\Omega|$ array $\mathcal{R}$, where $|\Omega| = \prod_{b=1}^{B} \binom{N_b}{n_b}$; however, that array is mostly not observed, and it is so large and detailed that it would be beyond human comprehension even if it were observed. We would like to define the treatment effect as a summary of $\mathcal{R}$, but in such a way that the summary is intelligible and usable in inference.

We define the treatment effect with reference to a uniformity trial of the type that, in a certain era, was commonly used as an aid to designing experiments; see Cochran (1937). For instance, uniformity trials were once used to study the performance of competing experimental designs, such as complete randomization or randomized blocks or randomized Latin squares. In a uniformity trial, treatment assignment $\mathbf{Z}$ is randomized as if an actual experiment were about to be performed, but instead $\mathbf{Z}$ is ignored and the standard treatment is applied in all cases. In its original use, a uniformity trial divides a farm into plots, assigns plots to a new treatment or a standard control at random, ignores the random assignment **Z** and applies in all cases the standard fertilizer, insecticide, etc., and ultimately harvests the crops recording yields in each plot. Essentially, the farmer cooperated in setting up an experiment and recording results, but he work the farm in the usual way, harvesting the usual crops for sale. This produced a simulated experiment with real crops in which the null hypothesis of no treatment effect is known to be true. For instance, by comparing two uniformity trials, perhaps at the same farm, one might discover that the estimated standard error is smaller from a uniformity trial designed as a Latin square than another uniformity trial designed as randomized blocks. In a certain era, statisticians did this, so in our era it is easy to imagine something that was once actually done.

We define the treatment effect with interference with reference to a uniformity trial. Stated informally, the effect a treatment with interference is a comparison of what happened in the actual experiment with its active treatment to what would have happened in a uniformity trial with the same treatment assignment **Z** but no active treatment. As in an era gone by, it is a comparison of two whole experiments, rather than a comparison of a treated and a control group. With interference, both treated and control units are affected by treatments applied to other units, so a comparison of a treated and a control group is not a comparison of a treated and an untreated situation. A comparison of an experiment with an active treatment and a uniformity trial is a comparison of a treated and an untreated situation. In §1.3, this is a comparison of an experiment with randomized stop and go trials and an experiment of identical structure with only go trials. Conveniently, using a few technical tools in §3.3 that were not available in 1937, we can make inferences about a uniformity trial that was never performed with the aid of a distribution-free pivotal quantity.

Write $\widetilde{r}_{bi}$ for the response of unit $i$ in block $b$ in the uniformity trial. There is only one such $\widetilde{r}_{bi}$, not one for each $\mathbf{z} \in \Omega$, because the realized treatment assignment $\mathbf{Z}$ that was recorded in an office has no way to affect the biological response of unit $i$ in block $b$. Because the uniformity trial was not actually performed, none of the $\widetilde{r}_{bi}$ are observed. Generally, $\widetilde{r}_{bi}$ need not equal any of $r_{bi\mathbf{z}}$, $\mathbf{z} \in \Omega$. If there were no interference between units, then $\widetilde{r}_{bi}$ would equal $r_{bi\mathbf{z}}$ for every $\mathbf{z} \in \Omega$ with $z_{bi} = 0$, because without interference the response of unit $bi$ depends only on the treatment $z_{bi}$ assigned to $bi$; however, with interference, it can happen that $\widetilde{r}_{bi} \neq r_{bi\mathbf{z}}$ for every $\mathbf{z} \in \Omega$. Write $\widetilde{\mathbf{r}} = (\widetilde{r}_{11}, \ldots, \widetilde{r}_{B,N_B})^T$.

In the presence of interference between units, the magnitude of the treatment effect is understood not as a comparison of treated and control groups both of which are affected by the treatment, but as a comparison of the actual experiment and the uniformity trial.

## 3 Inference with Interference

### 3.1 Preliminaries: a nonlinear rank statistic; testing no treatment effect

Fix an integer $k \geq 2$, with $k \leq \min_{b \in \{1,\ldots,B\}} m_b + 1$. As will be seen, the familiar choice is $k = 2$, and it yields the Mann-Whitney U-statistic, but there are reasons to prefer a larger value of $k$ when only some treated units respond to treatment. Ties among responses are not an issue in the fMRI experiment of §1.3, where blood oxygenation is recorded to many digits. We assume no ties in the discussion that follows.

The technical material that follows is not difficult but does require a certain amount of notation. To simplify, the reader may consider the special case of a single block ($B = 1$) with the parameter $k$ set to $k = 2$; then, one is considering a single-subject completely randomized trial, like the lady tasting tea, using the Mann-Whitney-Wilcoxon statistic, which happens to be the only linear placement statistic that is also a linear rank statistic (Orban and Wolfe 1982).

For a specific treatment assignment, $\mathbf{z} \in \Omega$, consider a subset $\mathcal{S}_{b\mathbf{z}} = \{i_1, \ldots, i_k\}$ of $k$ units from the same block $b$ with one treated unit, $z_{bi_1} = 1$, and $k-1$ control units, $z_{bi_j} = 0$, $j = 2, \ldots, k$. Write $\mathcal{K}_{b\mathbf{z}}$ for the collection of all $n_b \binom{m_b}{k-1}$ such subsets $\mathcal{S}_{b\mathbf{z}}$ for block $b$, so $\mathcal{S}_{b\mathbf{z}} \in \mathcal{K}_{b\mathbf{z}}$ if and only if $\mathcal{S}_{b\mathbf{z}} \subseteq \{1, \ldots, N_b\}$ with $|\mathcal{S}_{b\mathbf{z}}| = k$ and $1 = \sum_{i \in \mathcal{S}_{b\mathbf{z}}} z_{bi}$.

The set $\mathcal{S}_{b\mathbf{z}}$ compares one treated unit to $k-1$ control units. Write $v(\mathcal{S}_{b\mathbf{z}}) = 1$ if the treated unit, say $i_1$, in $\mathcal{S}_{b\mathbf{z}} = \{i_1, \ldots, i_k\}$ has the largest response under assignment $\mathbf{z}$, that is, if $r_{bi_1\mathbf{z}} > \max_{j \in \{i_2, \ldots, i_k\}} r_{bj\mathbf{z}}$, and write $v(\mathcal{S}_{b\mathbf{z}}) = 0$ otherwise. For $k = 2$, the set $\mathcal{S}_{b\mathbf{z}} = \{i_1, i_2\}$ has one treated unit and one control and $v(\mathcal{S}_{b\mathbf{z}}) = 1$ if the treated unit has a higher response than the control under assignment $\mathbf{z}$. Also, let $w_b$ be a weight to be attached to block $b$ where $w_b$ is a function of $\mathbf{n}$ and $\mathbf{m}$. In the current paper, $w_b = 1$ for all $b$, but another reasonable definition of $w_b$ will be given in a moment. For this treatment assignment $\mathbf{z}$, the quantity $T_\mathbf{z} = \sum_{b=1}^{B} w_b \sum_{\mathcal{S}_{b\mathbf{z}} \in \mathcal{K}_{b\mathbf{z}}} v(\mathcal{S}_{b\mathbf{z}})$ is a weighted count of the number of sets $\mathcal{S}_{b\mathbf{z}}$ such that the treated unit had a higher response than $k-1$ controls. With $w_b = 1$, the quantity $T_\mathbf{z}$ is a count, and a count is reasonable if the $N_b$ and $n_b$ do not vary much, as is true in §1.3. If the $n_b$ and $m_b$ varied greatly with $b$, then $w_b$ given by $1/w_b = B \, n_b \binom{m_b}{k-1}$ makes $T_\mathbf{z}$ the unweighted average over the $B$ blocks, $b = 1, \ldots, B$, of the proportion of sets $\mathcal{S}_{b\mathbf{z}} \in \mathcal{K}_{b\mathbf{z}}$ in which the treated unit had a higher response than $k-1$ controls, $v(\mathcal{S}_{b\mathbf{z}}) = 1$. For most $\mathbf{z}$, the quantity $T_\mathbf{z}$ depends upon parts of $\mathcal{R}$ that are not observed, so $T_\mathbf{z}$ cannot be computed from the observed data.

If the randomized treatment assignment $\mathbf{Z}$ replaces the specific treatment assignment $\mathbf{z}$, then needed parts of $\mathcal{R}$ are observed, and $T_\mathbf{Z}$ is a statistic that can be computed from the data. Indeed, if $B = 1$, $k = 2$, and $w_1 = 1$, then $T_\mathbf{Z}$ is the Mann-Whitney U-statistic and is linearly related to Wilcoxon's rank sum statistic. More generally, for $k = 2$ and $B > 2$, $T_\mathbf{Z}$ is a weighted sum of $B$ Mann-Whitney statistics; see Lehmann (1975, §3.3) and Puri (1965) who discuss weights intended to increase power against shift alternatives in the absence of interference.

For $k \geq 2$, if $n_b = 1$ and $N_b = N/B$ does not vary with $b$, then $T_\mathbf{Z}$ is the statistic discussed in Rosenbaum (2007b). Taking $k > 2$ tends to increase power when only a subset of treated units respond to treatment, as seems likely here for reasons discussed in §1.3. Indeed, with $k > 2$, the ranks are scored in a manner that closely approximates Conover and Salsburg's (1988) locally most powerful ranks for an alternative in which only a fraction of treated units respond, and the scores are identical to those proposed by Stephenson (1981).

In these special cases of the two previous paragraphs, $T_\mathbf{Z}$ is a stratified linear rank statistic. In general, $T_\mathbf{Z}$ is a function of the ranks, but not a linear function; however, it is a sum of $B$ linear functions of the placements within blocks in the sense of Orban and Wolfe (1982). For $B = 1$, the statistic with $k \geq 2$ has been discussed by Deshpande and Kochar (1980) and Stephenson and Gosh (1985) as an instance of Hoeffding's (1948) U-statistics under independent sampling of two distributions. Because interference precludes independent observations, inferences must be based on the random assignment of treatments, and for this the combinatorial development in Orban and Wolfe (1982) is particularly helpful.

Orban and Wolfe (1982) define the placement $m_b U_{bj}$ of the $j$th treated unit in block $b$ to be the number of controls in block $b$ who have a response less than or equal to the response of this treated unit. A linear placement statistic for one block $b$ is then of the form $\sum_{i=1}^{N_b} Z_{bi} \phi_{n_b,m_b}(U_{bj})$ for some function $\phi_{n_b,m_b}(\cdot)$. If there are no ties among responses within blocks, then taking $\phi_{n_b,m_b}(u) = w_b \binom{m_b u}{k-1}$ expresses $T_\mathbf{Z} = \sum_{b=1}^{B} w_b \sum_{\mathcal{S}_{b\mathbf{Z}} \in \mathcal{K}_{b\mathbf{z}}} v(\mathcal{S}_{b\mathbf{Z}})$ as a sum of linear functions of placements of the treated units, $T_\mathbf{Z} = \sum_{b=1}^{B} \sum_{j=1}^{n_b} \phi_{n_b,m_b}(U_{bj})$, where $\binom{\ell}{k-1}$ is defined to equal zero for $\ell < k - 1$.

Consider testing Fisher's null hypothesis $H_0$ of no effect which asserts that $r_{bi\mathbf{z}} = r_{bi\mathbf{z}'}$ for all $\mathbf{z}, \mathbf{z}' \in \Omega$ and all $b, i$. If $H_0$ were true, then the observed response $r_{bi\mathbf{Z}}$ is the same no matter what $\mathbf{Z} \in \Omega$ is randomly selected, so $\mathcal{R}$ is known and the distribution of $\Pr(T_{\mathbf{Z}} \geq t \mid \mathcal{R}, \mathbf{n}, \mathbf{m})$ is determined by the known fixed $\mathcal{R}$ and the randomized treatment assignment (1). Indeed, in part because no effect entails no interference in effects, testing no effect $H_0$ is a straightforward application of randomization inference. Orban and Wolfe (1982, §2) determine the null distribution of their linear placement statistic $\sum_{j=1}^{n_b} \phi_{n_b, m_b}(U_{bj})$ under independent sampling from a continuous distribution; however, their argument is entirely combinatorial, and it is easily seen that if responses with blocks are not tied then their argument and results give the exact null randomization distribution of $\sum_{j=1}^{n_b} \phi_{n_b, m_b}(U_{bj})$. Moreover, given $\mathcal{R}$, $\mathbf{n}$, $\mathbf{m}$, under $H_0$ and (1), $T_{\mathbf{Z}}$ is the sum of $B$ conditionally independent terms each with the known null distribution in Orban and Wolfe (1982, §2). Importantly, in the absence of ties, this null distribution of $T_{\mathbf{Z}}$ depends upon $\mathbf{n}$, $\mathbf{m}$, but not on $\mathcal{R}$.

### 3.2 The distribution of the test statistic in the uniformity trial

In the uniformity trial of §2.2, the null hypothesis of no effect on $\widetilde{r}_{bi}$ is known to be true because, following a concealed randomization $\mathbf{Z}$, no treatment was applied. Let $\widetilde{T}_{\mathbf{z}}$ be the value of the statistic of §3.1 computed from the $\widetilde{r}_{bi}$ in the uniformity trial when $\mathbf{Z} = \mathbf{z} \in \Omega$, with value $\widetilde{T}_{\mathbf{Z}}$ under the realized random assignment $\mathbf{Z}$. Specifically, write $\widetilde{v}(\mathcal{S}_{b\mathbf{z}}) = 1$ if the treated unit, say $i_1$, in $\mathcal{S}_{b\mathbf{z}} = \{i_1, \ldots, i_k\}$ has the largest response under assignment $\mathbf{z}$, that is, if $\widetilde{r}_{bi_1} > \max_{j \in \{i_2, \ldots, i_k\}} \widetilde{r}_{bj}$, and write $\widetilde{v}(\mathcal{S}_{b\mathbf{z}}) = 0$ otherwise, so that $\widetilde{T}_{\mathbf{z}} = \sum_{b=1}^{B} w_b \sum_{\mathcal{S}_{b\mathbf{z}} \in \mathcal{K}_{b\mathbf{z}}} \widetilde{v}(\mathcal{S}_{b\mathbf{z}})$. Even though $\widetilde{r}_{bi}$ is not affected by the treatment assignment $\mathbf{Z}$, the statistic $\widetilde{T}_{\mathbf{Z}}$ is generally a nondegenerate random variable because the value of the statistic depends jointly on the responses of units, $\widetilde{r}_{bi}$, which do not fluctuate, and on the treatments they receive, $Z_{bi}$, which are random.

In point of fact, neither $\widetilde{T}_{\mathbf{Z}}$ nor $\widetilde{T}_{\mathbf{z}}$ can be computed from observed data, because the uniformity trial was not performed and none of the $\widetilde{r}_{bi}$ are observed. Nonetheless, in the absence of ties, the distribution of $\widetilde{T}_{\mathbf{Z}}$ is known, because the null distribution in §3.1 depends upon $\mathbf{n}$ and $\mathbf{m}$ but not on $\mathcal{R}$; specifically, it is the convolution of $B$ random variables whose exact distributions are given by Orban and Wolfe (1982, Theorem 2.1, and expressions (2.1) and (2.2)), with expectation and variance

$$\mathrm{E}\left(\widetilde{T}_{\mathbf{Z}}\right) = \sum_{b=1}^{B} n_b \, \overline{\phi}_{n_b,m_b} \text{ where } \overline{\phi}_{n_b,m_b} = \frac{1}{m_b+1} \sum_{j=0}^{m_b} w_b \binom{j}{k-1},$$

$$\mathrm{var}\left(\widetilde{T}_{\mathbf{Z}}\right) = \sum_{b=1}^{B} \frac{n_b \left(n_b + m_b + 1\right)}{\left(m_b+1\right)\left(m_b+2\right)} \left[\left\{\sum_{j=0}^{m_b} w_b^2 \binom{j}{k-1}^2\right\} - (m_b+1)\,\overline{\phi}_{n_b,m_b}^2\right];$$

moreover, for reasonable choices of weights, $w_b$, as $B \to \infty$,

$$\left\{\widetilde{T}_{\mathbf{Z}} - \mathrm{E}\left(\widetilde{T}_{\mathbf{Z}}\right)\right\} / \sqrt{\mathrm{var}\left(\widetilde{T}_{\mathbf{Z}}\right)} \xrightarrow{D} \Phi\left(\cdot\right) \qquad (2)$$

where $\Phi\left(\cdot\right)$ is the standard Normal cumulative distribution. To emphasize, because the null hypothesis of no effect is known to be true in the uniformity experiment, and because the null distribution of $\widetilde{T}_{\mathbf{Z}}$ depends upon $\mathbf{n}$ and $\mathbf{m}$ but not on the $\widetilde{r}_{bi}$'s, it follows that we know the distribution of $\widetilde{T}_{\mathbf{Z}}$ in the uniformity trial even though we did not perform the uniformity trial and even if the treatment did have an effect with interference in the actual randomized experiment. This fact turns out to be useful with the aid of the concept of attributable effects (Rosenbaum 2001, 2007a, 2007b).

### 3.3 Attributable effects

Consider a specific treatment assignment $\mathbf{z} \in \Omega$ and the specific comparison $\mathcal{S}_{b\mathbf{z}} = \{i_1, \ldots, i_k\} \in \mathcal{K}_{b\mathbf{z}}$ of one treated unit, say $i_1$ with $z_{bi_1} = 1$, and $k-1$ control units, $i_j$ with $z_{bi_j} = 0$ for $j = 2, \ldots, k$. If $r_{bi_1\mathbf{z}} > \max_{j \in \{i_2, \ldots, i_k\}} r_{bj\mathbf{z}}$ then $i_1$ had the highest response in this compar-ison, contributing a 1 rather than a 0 to $T_\mathbf{z}$; however, this might or might not be an effect caused by the treatment, because even under the null hypothesis of no effect $H_0$, one of the $k$ units will have the highest response among the $k$ units. If $r_{bi_1\mathbf{z}} > \max_{j\in\{i_2,\ldots,i_k\}} r_{bj\mathbf{z}}$ but $\widetilde{r}_{bi_1} \leq \max_{j\in\{i_2,\ldots,i_k\}} \widetilde{r}_{bj}$ then treatment assignment $\mathbf{z}$ in the actual experiment does cause unit $i_1$ in block $b$ to have a higher response than units $\{i_2,\ldots,i_k\}$ in block $b$ in the sense that unit $i_1$ in block $b$ would not have had the highest response in this comparison in the uniformity trial of §2.2 in which no unit was treated. In §1.3, this would mean that in block $b$, stop trial $i_1$ caused activity in the STN region to exceed the level in go trials $i_2,\ldots,i_k$ in the sense that the activity was higher in the actual experiment and would not have been higher in the uniformity trial. Conversely, if $r_{bi_1\mathbf{z}} \leq \max_{j\in\{i_2,\ldots,i_k\}} r_{bj\mathbf{z}}$ but $\widetilde{r}_{bi_1} > \max_{j\in\{i_2,\ldots,i_k\}} \widetilde{r}_{bj}$ then treatment assignment $\mathbf{z}$ in the actual experiment prevented treated unit $i_1$ from having the highest response in $\mathcal{S}_{b\mathbf{z}}$, in the sense that $i_1$ would have had the highest response in the uniformity trial but did not have the highest response in the actual experiment. The third possibility is that treatment assignment $\mathbf{z}$ does not alter whether or not $i_1$ has the highest response in $\mathcal{S}_{b\mathbf{z}}$. Concisely, these three situations are: (i) $v(\mathcal{S}_{b\mathbf{z}}) = 1$ and $\widetilde{v}(\mathcal{S}_{b\mathbf{z}}) = 0$, (ii) $v(\mathcal{S}_{b\mathbf{z}}) = 0$ and $\widetilde{v}(\mathcal{S}_{b\mathbf{z}}) = 1$, and (iii) $v(\mathcal{S}_{b\mathbf{z}}) = \widetilde{v}(\mathcal{S}_{b\mathbf{z}})$.

For treatment assignment $\mathbf{z} \in \Omega$, the attributable effect

$$A_\mathbf{z} = T_\mathbf{z} - \widetilde{T}_\mathbf{z} = \sum_{b=1}^{B} w_b \sum_{\mathcal{S}_{b\mathbf{z}} \in \mathcal{K}_{b\mathbf{z}}} \{v(\mathcal{S}_{b\mathbf{z}}) - \widetilde{v}(\mathcal{S}_{b\mathbf{z}})\}$$

is the net increase in the number of times (weighted by $w_b$) that a treated response in the actual experiment exceeded $k-1$ control responses because of effects caused by using treatment assignment $\mathbf{z}$. So $A_\mathbf{z}$ is a real valued function of $\mathbf{z}$, $\widetilde{\mathbf{r}}$ and $\mathcal{R}$. In contrast, $A_\mathbf{Z}$ is the attributable effect for the $\mathbf{Z}$ randomly chosen according to (1), so $A_\mathbf{Z}$ is the difference between an observed statistic, $T_\mathbf{Z}$, that describes the actual experiment, and an unobservable random variable $\widetilde{T}_\mathbf{Z}$ that describes the uniformity trial in §2.2; however, the distribution of $\widetilde{T}_\mathbf{Z}$ is known, as discussed in §3.2. In brief, $A_\mathbf{Z}$ is an unknown random quantity which provides a reasonable measure of the effects of the treatment despite the presence of interference between units. More precisely, $A_\mathbf{Z}$ compares the aggregate effects of the treatment in the presence of interference to the pattern that would be exhibited in the uniformity trial in which no one is treated. If Fisher's null hypothesis of no effect $H_0$ is true, then $\mathrm{E}(A_\mathbf{Z}) = 0$. For discussion of attributable effects in randomized experiments without interference, see Rosenbaum (2001).

Let $\widetilde{t}_\alpha$ be the smallest value such that $\Pr\left(\widetilde{T}_\mathbf{Z} \leq \widetilde{t}_\alpha \mid \mathcal{R}, \mathbf{n}, \mathbf{m}\right) \geq 1 - \alpha$. From (2), for large $B$, we may approximate $\widetilde{t}_\alpha$ as

$$\widetilde{t}_\alpha \doteq \mathrm{E}\left(\widetilde{T}_\mathbf{Z}\right) + \Phi^{-1}(1-\alpha)\sqrt{\mathrm{var}\left(\widetilde{T}_\mathbf{Z}\right)}.$$

The following fact parallels a result in Rosenbaum (2007a) for a different family of statistics. In particular, Proposition 1 yields a one-sided $1-\alpha$ confidence interval for the unobserved random variable $A_\mathbf{Z}$ in terms of the observed random variable $T_\mathbf{Z}$ and the known quantity $\widetilde{t}_\alpha$. See Weiss (1955) for general discussion of confidence intervals for unobservable random variables.

**Proposition 1** *In a randomized experiment with interference in which (1) holds and there are no ties,*

$$\Pr\left(A_\mathbf{Z} \geq T_\mathbf{Z} - \widetilde{t}_\alpha \mid \mathcal{R}, \mathbf{n}, \mathbf{m}\right) \geq 1 - \alpha.$$

The proof of Proposition 1 is immediate:

$$\begin{aligned} \Pr\left(A_\mathbf{Z} \geq T_\mathbf{Z} - \widetilde{t}_\alpha \mid \mathcal{R}, \mathbf{n}, \mathbf{m}\right) &= \Pr\left(T_\mathbf{Z} - \widetilde{T}_\mathbf{Z} \geq T_\mathbf{Z} - \widetilde{t}_\alpha \mid \mathcal{R}, \mathbf{n}, \mathbf{m}\right) \\ &= \Pr\left(\widetilde{T}_\mathbf{Z} \leq \widetilde{t}_\alpha \mid \mathcal{R}, \mathbf{n}, \mathbf{m}\right) \geq 1 - \alpha. \end{aligned}$$

The attributable effect $A_\mathbf{Z}$ depends upon the sample sizes, $\mathbf{n}$ and $\mathbf{m}$, and the choice of $k$. Dividing $A_\mathbf{Z}$ by $\mathrm{E}\left(\widetilde{T}_\mathbf{Z}\right)$ can aid interpretation. Then $100 \times A_\mathbf{Z}/\mathrm{E}\left(\widetilde{T}_\mathbf{Z}\right)$ is the (weighted) percent increase above chance in the number of times a treated unit had a higher response than $k-1$ controls due to effects caused by the treatment, and with confidence $1-\alpha$, the unobserved $100 \times A_\mathbf{Z}/\mathrm{E}\left(\widetilde{T}_\mathbf{Z}\right)$ is at least $100 \times \left(T_\mathbf{Z} - \widetilde{t}_\alpha\right)/\mathrm{E}\left(\widetilde{T}_\mathbf{Z}\right)$.

We are suggesting that the unobservable random variable $A_\mathbf{Z}/\mathrm{E}\left(\widetilde{T}_\mathbf{Z}\right)$ is a useful measure of the magnitude of the treatment effect when interference may be present; however, it is a new measure, and its magnitude is unfamiliar. To build some intuition about magnitudes of $A_\mathbf{Z}/\mathrm{E}\left(\widetilde{T}_\mathbf{Z}\right)$, consider its behavior in a familiar context, namely a single block, $B=1$, independent observations without interference and a treatment effect that is an additive shift, $\delta$. In this case, as $n_1 \to \infty$ and $m_1 \to \infty$, the quantity $T_\mathbf{Z}/\left\{n_1\binom{m_1}{k-1}\right\}$ converges in probability to the probability that a treated response exceeds $k-1$ control responses and $A_\mathbf{Z}/\mathrm{E}\left(\widetilde{T}_\mathbf{Z}\right)$ converges in probability the percent increase in this probability above the chance level of $1/k$. Table 1 evaluates these limits for the standard Normal distribution and the $t$-distribution with 2 degrees of freedom. For instance, with a shift $\delta$ in a Normal that equals a full standard deviation, $\delta = 1$, the probability that a treated response exceeds nine control responses in 0.341 which is 241% above the chance level of 0.1 for $\delta = 1$. The quantity $A_\mathbf{Z}/\mathrm{E}\left(\widetilde{T}_\mathbf{Z}\right)$ has the advantage that it continues to be meaningful with interference where shift models are inapplicable.

Proposition 1 refers to an analysis of responses, but it is possible in a randomized experiment to use the same approach with residuals from a robust covariance adjustment which controls for measured disturbing covariates such as head motion. See Rosenbaum (2002) for general discussion of randomization inference for covariance adjustment in randomized experiments, and see Rosenbaum (2007a, §6) for its application with interference. This procedure is illustrated in §4 in Table 4.

Table 1: In the absence of interference and dependence, the upper table gives the probability that one treated response is higher than $k-1$ independent control responses when the treatment effect is an additive shift $\delta$ and the errors are independently drawn from either a standard Normal distribution or the $t$-distribution with 2 degrees of freedom. The lower table gives the percentage increase in this probability above chance; for example, $14\% = (0.57 - 0.50)/0.50$.

|  | Probability a treated response is higher than $k-1$ controls | | | | | |
|---|---|---|---|---|---|---|
| $\delta$ | Normal | | | $t$ with 2 df | | |
|  | $k=2$ | $k=5$ | $k=10$ | $k=2$ | $k=5$ | $k=10$ |
| 0 | 0.50 | 0.20 | 0.10 | 0.50 | 0.20 | 0.10 |
| 0.25 | 0.57 | 0.26 | 0.14 | 0.55 | 0.24 | 0.12 |
| 0.5 | 0.64 | 0.33 | 0.20 | 0.60 | 0.29 | 0.15 |
| 1 | 0.76 | 0.49 | 0.34 | 0.69 | 0.39 | 0.22 |

|  | Percentage increase above chance | | | | | |
|---|---|---|---|---|---|---|
| $\delta$ | Normal | | | $t$ with 2 df | | |
|  | $k=2$ | $k=5$ | $k=10$ | $k=2$ | $k=5$ | $k=10$ |
| 0 | 0 | 0 | 0 | 0 | 0 | 0 |
| 0.25 | 14 | 31 | 44 | 10 | 21 | 22 |
| 0.5 | 28 | 67 | 99 | 20 | 45 | 50 |
| 1 | 52 | 147 | 241 | 39 | 97 | 120 |

## 4  To what extent do stop trials activate the subthalamic nucleus?

Is activity in the subthalamic nucleus (STN) elevated following stop trials? Figures 1 and 2 depict STN activity for one subject in one session, but there are 58 subjects, each with 4 sessions, making $58 \times 4 = 232$ blocks, with a total of $N = 22,440$ randomized go-or-stop trials.

Table 2 performs the analysis in §3 three times, for $k = 2$, 5 and 10. Recall that for $k = 2$, the statistic $T_{\mathbf{Z}}$ is the sum of 232 Mann-Whitney-Wilcoxon statistics. The deviates for testing the null hypothesis of no effect $H_0$ are extremely large, particularly for the filtered data with $k = 5$ or $k = 10$. In the uniformity trial, we expect that when comparing a treated unit to nine controls, one time in ten the treated unit will have the highest response. For filtered STN, $k = 10$, the point estimate of $A_{\mathbf{Z}}/\mathrm{E}\left(\widetilde{T}_{\mathbf{Z}}\right)$ suggests a 53.0% increase above this chance expectation due to effects caused by the treatment, but we are 95% confident of only a 46.4% increase. Again, in the presence of arbitrary interference between units, these are correct statements about the relationship between the actual trial, with its unobserved attributable effect $A_{\mathbf{Z}}$, and the uniformity trial that was not actually performed.

In addition to the subthalamic nucleus, other regions of the brain are suspected to be involved in motor response inhibition, including the right inferior frontal cortex (or rIFC, see Fortsmann, et al. 2008) and the presupplementary motor area (or preSMA, see Simmonds, Pekar and Mostofsky 2008). In analyses parallel to Table 2, we found smaller but significantly elevated activity in both the rIFC and preSMA. Using filtered data for rIFC with $k = 5$, we obtained a $P$-value testing no effect of 0.0030, a point estimate for $A_{\mathbf{Z}}/\mathrm{E}\left(\widetilde{T}_{\mathbf{Z}}\right)$ of 0.059 and a 95% confidence interval of $A_{\mathbf{Z}}/\mathrm{E}\left(\widetilde{T}_{\mathbf{Z}}\right) \geq 0.024$. Using filtered data for preSMA with $k = 5$, we obtained a $P$-value testing no effect of 0.000053, a point estimate for $A_{\mathbf{Z}}/\mathrm{E}\left(\widetilde{T}_{\mathbf{Z}}\right)$ of 0.084 and a 95% confidence interval of $A_{\mathbf{Z}}/\mathrm{E}\left(\widetilde{T}_{\mathbf{Z}}\right) \geq 0.048$.

Table 2: Randomization test of no treatment effect $H_0$ and randomization-based confidence interval for the attributable effect $A_\mathbf{Z}$ in the presence of interference between units.

| | Test of No Effect | Fractional Increase $A_\mathbf{Z}/\mathrm{E}\left(\widetilde{T}_\mathbf{Z}\right)$ | |
|---|---|---|---|
| | Deviate Testing $H_0$ $\frac{(T_\mathbf{Z}-\mathrm{E}(\widetilde{T}_\mathbf{Z}))}{\sqrt{\mathrm{var}(\widetilde{T}_\mathbf{Z})}}$ | Point Estimate $\frac{T_\mathbf{Z}-\mathrm{E}(\widetilde{T}_\mathbf{Z})}{\mathrm{E}(\widetilde{T}_\mathbf{Z})}$ | 95% CI $\frac{T_\mathbf{Z}-\widetilde{t}_\alpha}{\mathrm{E}(\widetilde{T}_\mathbf{Z})}$ |
| $k$ | Unfiltered STN | | |
| $k=2$ | 8.427 | 0.076 | 0.061 |
| $k=5$ | 8.874 | 0.192 | 0.156 |
| $k=10$ | 8.161 | 0.327 | 0.261 |
| | Filtered STN | | |
| $k=2$ | 11.000 | 0.099 | 0.084 |
| $k=5$ | 13.630 | 0.295 | 0.259 |
| $k=10$ | 13.219 | 0.530 | 0.464 |

Although the point estimates of 5.9% above chance for rIFC and 8.4% above chance for preSMA are significantly larger than zero, they are substantially smaller than the point estimate of 29.5% above chance for filtered STN with $k=5$ in Table 2.

For $k > 2$, the statistic $T_\mathbf{Z}$ and unmeasurable attributable effect $A_\mathbf{Z}$ handle the treated and control groups in different ways: one treated unit is compared to $k-1$ controls. If one expected successful stop trials to suppress rather than elevate activity, one needs to apply $T_\mathbf{Z}$ to $-r_{bi\mathbf{Z}}$ rather than to $r_{bi\mathbf{Z}}$. For instance, we might expect reduced activity in the primary motor cortex (M1) during stop trials, because motor activity is not requested in a stop trial. Applying $T_\mathbf{Z}$ to $-r_{bi\mathbf{Z}}$ for filtered data from M1 with $k=5$, we obtain a $P$-value testing no effect of 0.000011, a point estimate for $A_\mathbf{Z}/\mathrm{E}\left(\widetilde{T}_\mathbf{Z}\right)$ of 0.092 and a 95% confidence interval of $A_\mathbf{Z}/\mathrm{E}\left(\widetilde{T}_\mathbf{Z}\right) \geq 0.056$, consistent with reduced activity in M1 in stop trials.

The inferences just described are appropriate in the presence of interference of arbitrary form. But is there interference? Here, we look at one very simple possible form for interference, namely interference from the immediately previous trial. Recall that trials are randomly go or stop trials, where go trials occur with probability .75 and stop trials with probability 0.25. Aside from the first trial in a session, the remaining trials may be classified into four groups based on the current and previous trial as go-go with probability $0.75^2 = 0.5625$, stop-go with probability $0.25 \times 0.75 = 0.1875$, go-stop with probability $0.75 \times 0.25 = 0.1875$, and stop-stop with probability $0.25^2 = 0.0625$. If there is no interference, then the treatment at the current trial may have an effect, but not the treatment at the previous trial, so go-go should have the same effect as stop-go, and go-stop should have the same effect as stop-stop. Table 3 compares the two common cases, go-go to stop-go trials, ignoring other cases, using the same methods as in Table 2, reporting only results for filtered data. In Table 3, a difference indicates a very specific form of interference, namely a lingering effect from a previous stop trial on a current go trial. There is clearly evidence in Table 3 of a lingering effect of a previous stop trial, but the magnitudes of effect are much smaller than in Table 2 for the effect of the treatment in the current trial. Importantly, the inferences in Table 3 are appropriate for comparing go-go and stop-go trials even if other forms of interference are also present.

Table 3: Testing for a simple form of interference: comparison of go-go trials and stop-go trials for STN.

|         | Test of No Lingering Effect | Fractional Increase $A_{\mathbf{Z}}/\mathrm{E}\left(\widetilde{T}_{\mathbf{Z}}\right)$ | |
|---------|:---:|:---:|:---:|
|         | Deviate | Point Estimate | 95% CI |
|         | Filtered | | |
| $k = 2$ | 2.906 | 0.031 | 0.013 |
| $k = 5$ | 3.151 | 0.085 | 0.040 |
| $k = 10$ | 3.095 | 0.201 | 0.094 |

Head movements during the experiment may distort fMRI readings. As discussed and illustrated in Rosenbaum (2007, §6), instead of studying the attributable effect for the responses themselves, the method in Rosenbaum (2002) may be used as the basis for randomization inference about the attributable effect on residuals from a robust covariance adjustment. Table 4 applies the method of Table 2 to residuals of STN levels after adjustment for six covariates describing translation and rotation of the head as estimated from three-dimensional images of each session, the residuals being obtaining using the default settings of the R function rlm which implements Huber's m-estimation. Table 4 is generally similar to Table 2, so covariance adjustment for head motion did not greatly alter the results.

Table 4: Comparison of STN activity with robust covariance adjustment for head movement.

|  | Test of No Effect $H_0$ | Fractional Increase $A_{\mathbf{Z}}/\mathrm{E}\left(\widetilde{T}_{\mathbf{Z}}\right)$ | |
|---|---|---|---|
|  | Deviate | Point Estimate | 95% CI |
|  | Filtered | | |
| k=2 | 10.974 | 0.099 | 0.084 |
| k=5 | 12.528 | 0.271 | 0.235 |
| k=10 | 11.103 | 0.445 | 0.379 |

## 5 A Simulation of the Size and Power of Competing Tests in the Presence of Interference Between Units

### 5.1 Description of the simulation

Tables 5 and 6 report a simulation study of power with and without interference between units. Both tables refer to a completely randomized experiment; that is, there is a single block, $B = 1$. In Table 5, there are $N = 250$ trials, whereas in Table 6 there are $N = 1000$ trials. Each trial is randomly assigned to be a treated trial or a control trial with probability $\frac{1}{2}$. As in the actual experiment in §1.3, only some treated trials elicit the intended cognitive activity and brain response. In Table 5, $\lambda = 50\%$ of treated trials are successful, whereas in Table 6, $\lambda = 10\%$ of treated trials are successful. A control trial yields a response drawn from a distribution $F(\cdot)$, and in Tables 5 and 6 this distribution $F(\cdot)$ is either the standard Normal distribution or the $t$-distribution on 2 degrees of freedom. In the absence of interference, a successful treated trial yields a response from $F^\nu(\cdot)$ and an unsuccessful trial yields a response from $F(\cdot)$, so an unsuccessful treated trial looks like a control trial, but a successful treated trial looks like the maximum of $\nu$ independent control trials; see Lehmann (1953), Salzburg (1986) and Conover and Salzburg (1988) for discussion and history of this mixture model. Formally, in the absence of interference, the Salzburg model yields control responses from $F(\cdot)$ and treated responses from $(1-\lambda)F(\cdot)+\lambda F^\nu(\cdot)$, where Lehmann had considered $\nu = 2$, and Conover and Salzburg had determined the locally most powerful ranks as $\lambda \to 0$, which are essentially Wilcoxon's ranks for $\nu = 2$. In this mixture model, successful treated trials are from $F^\nu(\cdot)$ and unsuccessful treated trials are from $F(\cdot)$, but trials are not labeled as successful or unsuccessful. We introduce interference into this model by assuming that a successful trial samples from $F^\nu(\cdot)$ rather than $F(\cdot)$ only if certain additional conditions hold defined in terms of treatments assigned to the previous few trials.

In Table 5, $\nu = 10$, but in Table 6, $\nu = 20$; that is, a larger $\nu$ in Table 6 offsets a smaller $\lambda$ so that the power remains in an interesting range. The maximum of $\nu = 10$ independent observations from a Normal distribution will often be smaller and more stable than the maximum of $\nu = 10$ observations from a $t$-distribution with 2 degrees of freedom, and this may affect different tests in different ways.

Four types of interference were simulated. With interference of type A, a successful treated trial that immediately follows a control trial has a response drawn from $F^\nu(\cdot)$, but all unsuccessful trials and all treated trials that immediately follow other treated trials have responses drawn from $F(\cdot)$. With interference of type B, a successful treated trial that immediately follows a treated trial has a response drawn from $F^\nu(\cdot)$, but all unsuccessful trials and all treated trials that immediately follow control trials have responses drawn from $F(\cdot)$. Interference C and D resemble interference A, except that in C a successful treated trial only has a response drawn from $F^{\nu}(\cdot)$ if it follows 2 or more control trials, and in D if it follows 3 or more control trials.

Table 5: Simulated power with interference in a randomized experiment in a single block, $B = 1$, of size $N = 250$, when 50% of trials are successful, $\lambda = 0.5$. The case $\nu = 1$ is the null hypothesis of no effect and hence no interference among effects, so the simulation is estimating the true size of a test with nominal level 0.05. The statistic $k = 2$ is the Mann-Whitney-Wilcoxon statistic. The highest power in a non-null row is in **bold**.

|  | $\lambda = 0.5$, $N = 250$ | | | |
|---|---|---|---|---|
|  | No Autoregressive Errors Added | | | |
|  | $F(.)$ is Normal | | | |
|  | t-test | $k = 2$ | $k = 5$ | $k = 10$ |
| $\nu = 1$, No effect | 0.0434 | 0.0462 | 0.0432 | 0.0412 |
| $\nu = 10$, No interference | 0.9992 | 0.9998 | **1.0000** | 0.9856 |
| $\nu = 10$, Interference A | 0.8028 | 0.8014 | **0.8928** | 0.7328 |
| $\nu = 10$, Interference B | 0.8006 | 0.7968 | **0.8810** | 0.7274 |
| $\nu = 10$, Interference C | 0.3056 | 0.2830 | **0.3704** | 0.2806 |
| $\nu = 10$, Interference D | 0.1174 | 0.1060 | **0.1238** | 0.0896 |
|  | $F(.)$ is the t-distribution, 2 df | | | |
|  | t-test | $k = 2$ | $k = 5$ | $k = 10$ |
| $\nu = 1$, No effect | 0.0410 | 0.0448 | 0.0430 | 0.0436 |
| $\nu = 10$, No interference | 0.9542 | **1.0000** | **1.0000** | 0.9854 |
| $\nu = 10$, Interference A | 0.6610 | 0.8130 | **0.9004** | 0.7392 |
| $\nu = 10$, Interference B | 0.6510 | 0.7998 | **0.8892** | 0.7316 |
| $\nu = 10$, Interference C | 0.2464 | 0.2838 | **0.3652** | 0.2704 |
| $\nu = 10$, Interference D | 0.0966 | 0.1188 | **0.1302** | 0.0984 |
|  | Autoregressive Errors Added | | | |
|  | $F(.)$ is Normal | | | |
|  | t-test | $k = 2$ | $k = 5$ | $k = 10$ |
| $\nu = 1$, No effect | 0.0494 | 0.0518 | 0.0458 | 0.0476 |
| $\nu = 10$, No interference | 0.9714 | **0.9744** | 0.9454 | 0.7670 |
| $\nu = 10$, Interference A | **0.4868** | 0.4824 | 0.4528 | 0.2976 |
| $\nu = 10$, Interference B | **0.4874** | 0.4772 | 0.4572 | 0.3002 |
| $\nu = 10$, Interference C | **0.1622** | 0.1562 | 0.1498 | 0.0892 |
| $\nu = 10$, Interference D | **0.0786** | 0.0746 | 0.0726 | 0.0524 |
|  | $F(.)$ is the t-distribution, 2 df | | | |
|  | t-test | $k = 2$ | $k = 5$ | $k = 10$ |
| $\nu = 1$, No effect | 0.0442 | 0.0524 | 0.0476 | 0.0476 |
| $\nu = 10$, No interference | 0.9506 | 0.9968 | **0.9976** | 0.9670 |
| $\nu = 10$, Interference A | 0.5826 | 0.6602 | **0.7494** | 0.6184 |
| $\nu = 10$, Interference B | 0.5810 | 0.6534 | **0.7316** | 0.5926 |
| $\nu = 10$, Interference C | 0.1996 | 0.2218 | **0.2502** | 0.1826 |
| $\nu = 10$, Interference D | 0.0866 | 0.0928 | **0.0950** | 0.0704 |

Table 6: Simulated power with interference in a randomized experiment in a single block, $B = 1$, of size $N = 1000$, when 10% of trials are successful, $\lambda = 0.1$. The case $\nu = 1$ is the null hypothesis of no effect and hence no interference among effects, so the simulation is estimating the true size of a test with nominal level 0.05. The statistic $k = 2$ is the Mann-Whitney-Wilcoxon statistic.

|  | $\lambda = 0.1$, $N = 1000$ | | | |
|---|---|---|---|---|
|  | No Autoregressive Errors Added | | | |
|  | $F(.)$ is Normal | | | |
|  | t-test | $k = 2$ | $k = 5$ | $k = 10$ |
| $\nu = 1$, No effect | 0.0506 | 0.0518 | 0.0484 | 0.0498 |
| $\nu = 20$, No interference | 0.8020 | 0.7050 | 0.9416 | **0.9710** |
| $\nu = 20$, Interference A | 0.3044 | 0.2366 | 0.4294 | **0.5052** |
| $\nu = 20$, Interference B | 0.3110 | 0.2466 | 0.4266 | **0.5052** |
| $\nu = 20$, Interference C | 0.1094 | 0.0914 | 0.1414 | **0.1598** |
| $\nu = 20$, Interference D | 0.0672 | 0.0652 | **0.0714** | 0.0654 |
|  | $F(.)$ is the t-distribution, 2 df | | | |
|  | t-test | $k = 2$ | $k = 5$ | $k = 10$ |
| $\nu = 1$, No effect | 0.0452 | 0.0544 | 0.0506 | 0.0500 |
| $\nu = 20$, No interference | 0.6992 | 0.6938 | 0.9316 | **0.9658** |
| $\nu = 20$, Interference A | 0.2788 | 0.2476 | 0.4292 | **0.4996** |
| $\nu = 20$, Interference B | 0.2624 | 0.2354 | 0.4278 | **0.4956** |
| $\nu = 20$, Interference C | 0.1024 | 0.0934 | 0.1364 | **0.1576** |
| $\nu = 20$, Interference D | 0.0636 | 0.0646 | 0.0698 | **0.0712** |
|  | Autoregressive Errors Added | | | |
|  | $F(.)$ is Normal | | | |
|  | t-test | $k = 2$ | $k = 5$ | $k = 10$ |
| $\nu = 1$, No effect | 0.0480 | 0.0486 | 0.0474 | 0.0434 |
| $\nu = 20$, No interference | 0.4634 | 0.4254 | **0.5342** | 0.5034 |
| $\nu = 20$, Interference A | 0.1600 | 0.1472 | **0.1810** | 0.1604 |
| $\nu = 20$, Interference B | 0.1646 | 0.1494 | **0.1852** | 0.1672 |
| $\nu = 20$, Interference C | 0.0760 | 0.0736 | **0.0764** | 0.0714 |
| $\nu = 20$, Interference D | 0.0500 | 0.0496 | **0.0538** | 0.0516 |
|  | $F(.)$ is the t-distribution, 2 df | | | |
|  | t-test | $k = 2$ | $k = 5$ | $k = 10$ |
| $\nu = 1$, No effect | 0.0450 | 0.0474 | 0.0460 | 0.0486 |
| $\nu = 20$, No interference | 0.6610 | 0.6270 | 0.8600 | **0.9030** |
| $\nu = 20$, Interference A | 0.2462 | 0.2052 | 0.3372 | **0.3938** |
| $\nu = 20$, Interference B | 0.2408 | 0.2000 | 0.3398 | **0.3914** |
| $\nu = 20$, Interference C | 0.0978 | 0.0828 | 0.1128 | **0.1274** |
| $\nu = 20$, Interference D | 0.0620 | 0.0608 | 0.0648 | **0.0658** |

Interference between units creates one type of dependence over successive trials, but there can also be other types of dependence that are present in the absence of interference, indeed present in the absence of any treatment effect. The upper half of Tables 5 and 6 is dependent over successive trials only due to interference. In the lower half of Tables 5 and 6, the responses above are added to stationary autoregressive errors with standard Normal marginal distributions and autocorrelation 0.5.

Each situation was simulated 5000 times, so the simulated power has a standard error of at most $\sqrt{.25/5000} = 0.007$.

## 5.2 Results of the simulation

Tables 5 and 6 contrast the size and power of four test statistics, namely the conventional pooled variance $t$-statistic and $T_{\mathbf{Z}}$ for three values of $k$, $k = 2$, $k = 5$, and $k = 10$. Recall that $k = 2$ corresponds with the Mann-Whitney-Wilcoxon statistic, and $k = 5$ is similar to the suggestion of Salzburg (1986) and Conover and Salzburg (1988).

The case of $\nu = 1$ in Tables 5 and 6 is the null hypothesis: it suggests that all four tests have size close to their nominal level of 0.05 in all sampling situations. This is expected for $T_{\mathbf{Z}}$ because it is a randomization test applied under the null hypothesis of no effect in a randomized experiment. For related results about the randomization distribution of statistics such as the $t$-statistic, see Welch (1937). Notice that, because this is a randomization test in a randomized experiment, it has the correct level even in the case of autocorrelated errors. In brief, because all four tests appear to be valid, falsely rejecting true hypotheses at the nominal rate of 5%, it is reasonable to contrast the tests in terms of power.

In the non-null cases, $\nu > 1$, the test with the highest power is in **bold**. No one test is uniformly best in the situations considered in Tables 5 and 6, but the t-test and the Mann-Whitney-Wilcoxon test are never much better than $k = 5$ and are often much worse. The statistic with $k = 10$ performs well only in Table 6 where successful trials occur only 10% of the time. The permutational $t$-statistic performs well only when both $F(\cdot)$ and the autoregressive errors are Normal, and it performs poorly when the $F(\cdot)$ is the $t$-distribution with 2 df. When $F(\cdot)$ is Normal and there are no autoregressive errors, the permutational $t$-statistic typically had less power than $k = 5$.

Tables 5 and 6 exhibit many patterns. It is not surprising that the addition of Gaussian autoregressive errors reduces power: the power in the top half of Tables 5 and 6 is typically quite a bit higher than the corresponding power in the lower half of the tables. Both types of interference, A and B, reduce power when compared with no interference, but in Tables 5 and 6 interference A and B had similar effects on power. Interference patterns C and D reduce the number of responses that differ from control, so they reduce power relative to case A, but the suggestion of Conover and Salsburg, namely $k = 5$, exhibits decent relative performance in most if not all cases.

### 5.3 Comparison with SPM

A common approach to the analysis of fMRI data is the statistical parametric map (SPM) approach of Friston et al. (1995). Using responses convolved with a hemodynamic response function (HRF) as in Figure 2, the SPM approach entails testing a hypothesis of the equality of two regression coefficients in a generalized least squares analysis. We simulated this analysis with and without interference, with and without autoregressive errors, and with mixtures of successful and unsuccessful trials. The SPM approach uses a parametric model and is not a randomization inference, so there is no reason to expect that it will have the correct level when there is no treatment effect but the parametric model is false. Indeed, in nominal 0.05 level tests in our simulation, true null hypotheses of no effect were rejected more often than 5% of the time, in some cases with probabilities as high as 30%. In light of this, a power comparison is not appropriate. It is not a fault of the SPM method that it does not control the type one error rate when the null hypothesis of no effect is true but the model itself is false; that type of error control is not expected from standard parametric inference. Presumably, a careful user of the SPM approach would check for model failures using residuals and diagnostics, and alter the parametric model in appropriate ways. Nonetheless, it is convenient that the randomization inferences in §3 do control the type one error rate at 5% in the presence of autocorrelation, interference between units, unsuccessful trials and error distributions (such as the $t$-distribution with 2 degrees of freedom) that lack a finite variance.

## 5.4 Alternative designs and power

The simulation has compared the power of different statistics in given situations with interference. Another potential source of increased power entails alternative experimental designs which alter the degree of interference by altering the time interval between trials. In the absence of interference, we generally expect more power with more trials, so naively we might expect increased power from ever more trials ever more rapidly paced. However, in cross-over designs, it is also commonly said that interference should be reduced by allowing time for a wash-out period between trials. In particular, it is possible that few trials with more time between them would yield less interference and fMRI activity that is more sharply distinct following treatment or control. If one were using a statistic that is valid only in the absence of interference, then the power in these two situation could not be compared, because a broader range of validity is being weighed against possibly reduced power. In contrast, using the randomization distribution of $T_{\mathbf{Z}}$ to test $H_0$, the test is valid, with correct level, for both rapid-fire designs with many trials and widely-spaced designs with fewer trials, and a comparison of power is possible. For instance, a smaller number of trials with more successful trials and less interference ($N = 250$, $\lambda = .5$, case A in Table 3) yields greater power than more trials with fewer successful trials and more interference ($N = 1000$, $\lambda = .1$, case D in Table 4), so it is clear that increasing the number of trials must be weighed against potential harms from increasing the pace at which trials are conducted.

## 6 Summary

Randomized experiments in cognitive neuroscience of the type described in §1.3 have three attributes that were important in the current discussion. First, with about 100 randomized stimuli for a single brain in a session of 600 seconds, interference is likely: the stimulus applied in one trial is likely to affect the response measured for other trials. Interference that is local in time is almost inevitable because the measurable response to one stimulus lasts for more than six seconds, but additionally as the trial progresses a subject is growing more familiar and experienced with the tasks and equipment, so interference may have a complex form that can extend across different sessions for the same subject. The use of the HRF function in passing from Figure 1 to Figure 2 is a standard attempt to pick out the response to a particular stimulus, and useful though this is, it is at best an approximation. Second, with rapid fire trials of this sort, not every trial will be successful in eliciting the intended cognitive activity. This is quite evident in the experiment in §1.3, because subjects responded inappropriately to some go or stop trials, but inattention, distraction or confusion can also occur without visible evidence. Some exposures to a stimulus stimulate the intended thought process, some don't. Third, because this is a randomized experiment, randomization can form the basis for inference, thereby avoiding assumptions of independence and non-interference. Within this context, we have proposed and illustrated a straightforward, robust methodology that (i) yields a confidence interval for the magnitude of effect despite interference between units, and (ii) often has greater power than procedures based on Wilcoxon's statistic when only some treated trials are successful.

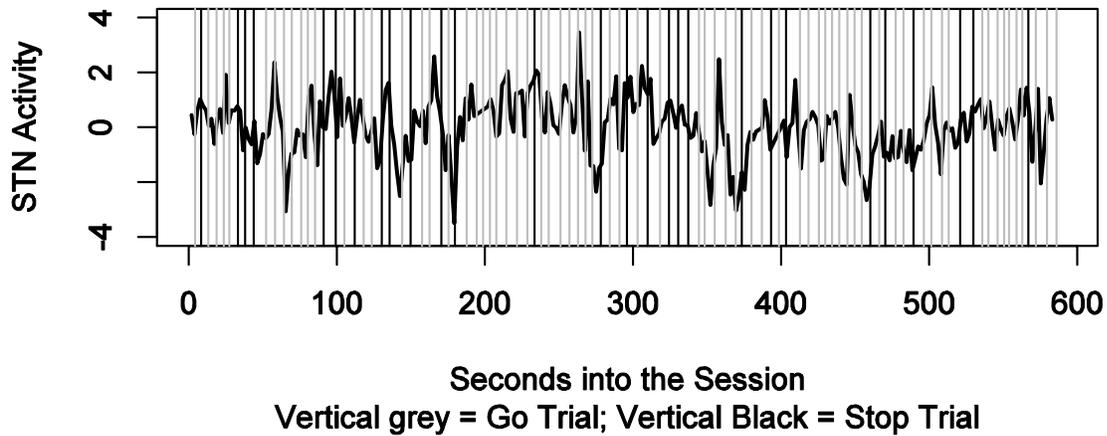

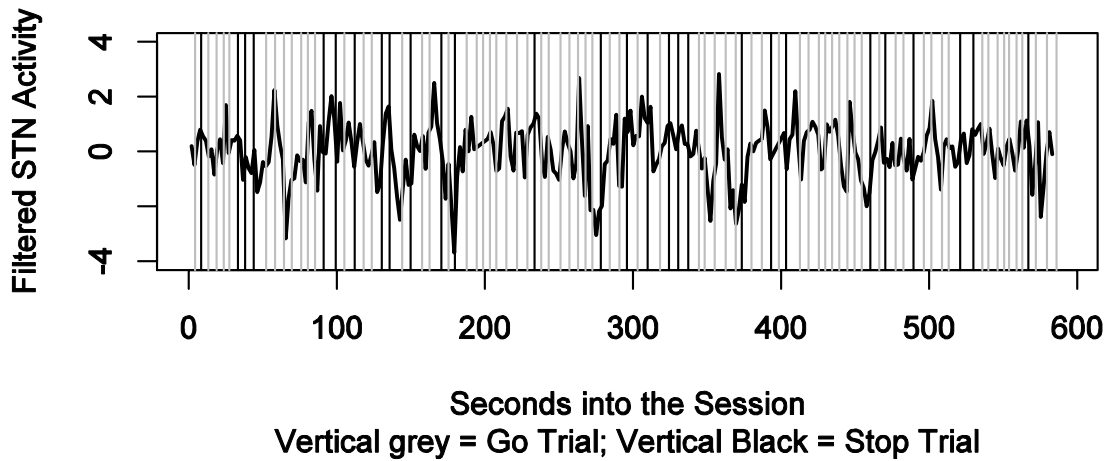

Figure 1: One session of the experiment for one subject. Unfiltered and filtered activity in the subthalamic nucleus is depicted every two seconds for roughly ten minutes. With probability ¾, the next trial is a go trial (grey) and with probability ¼ it is a stop trial (black).

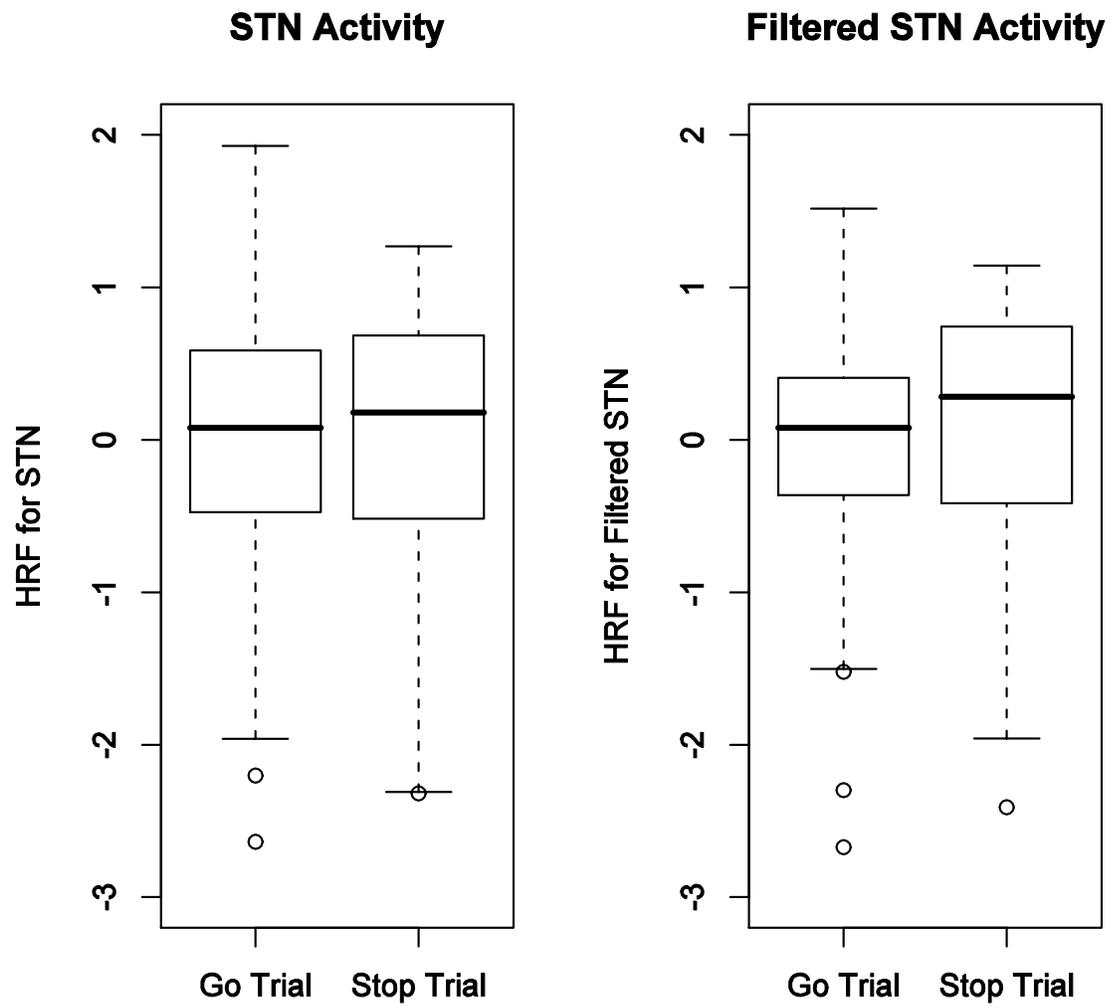

Figure 2. Hemodynamic response function (HRF) for the subthalamic nucleus after each trial for one subject in one session.